\documentclass[10pt]{article} 

\usepackage{geometry} 

\usepackage{fancyhdr}
\usepackage{extramarks}
\usepackage{lastpage}

\fancypagestyle{plain} {
\renewcommand{\headrulewidth}{0pt}
\renewcommand{\footrulewidth}{0pt}
\fancyhf{}
\headsep=0.8cm
\fancyhead[C]{74th International Astronautical Congress (IAC), Baku, Azerbaijan, 2-6 October 2023. \\ Copyright ©2023 by the International Astronautical Federation (IAF). All rights reserved.}
\footskip=1.5cm
\fancyfoot[L]{IAC-23-\conftitle}
\fancyfoot[R]{Page \thepage\ of \pageref{LastPage}}
}
\usepackage{multicol} 

\usepackage{authblk}
\newcommand{\authorIAC}[2]{\author[#1]{ {\uppercase{\bf\normalsize #2}}}}
\newcommand{\affiliation}[2]{ \affil[#1 ]{\raggedright{\textit{\normalsize #2}}}}
\newcommand{\email}[1]{\affil[* ]{\normalsize #1}}

\usepackage{titlesec}
\titleformat{\section}
    {\normalfont\normalsize\bfseries}{\thesection.}{0.5em}{}
    
\titleformat{\subsection}
    {\normalfont\normalsize\it}{\thesubsection.}{0.5em}{}
    
\titleformat{\subsubsection}
    {\normalfont\normalsize\it}{\thesubsubsection.}{0.5em}{}

\titlespacing{\section}{0pt}{0.8cm}{1pt}
\titlespacing{\subsection}{0pt}{0.7cm}{1pt}
\titlespacing{\subsubsection}{0pt}{0.6cm}{1pt}

\usepackage{mathptmx}
\usepackage[scaled=.90]{helvet}
\usepackage{courier}

\usepackage[utf8]{inputenc}
\usepackage[hidelinks]{hyperref}

\usepackage{ctable}

\usepackage{subfiles}
\usepackage{float}
\usepackage{amsmath}
\usepackage{graphicx}
\usepackage{booktabs}
\usepackage{caption}
\usepackage{subcaption}
\usepackage{physics}
\usepackage{array}
\usepackage{color,soul}

\def \be {\begin{equation}}
\def \ee {\end{equation}}

\def \O {${\rm O^+}$}

\usepackage{blindtext}

\def \conftitle {\hl{B4.IP.6}}

\title{ {\normalsize IAC-23-\conftitle \\ \vspace{4mm}

{\bf Spacecraft Charging of the Morazán MRZ-SAT Satellite in Low Earth Orbit:

Initial Results on the Influence of Energetic Electron Anisotropy on Differential Charging}}}

\authorIAC{a b *}{R. Bertrand-Delgado}

\authorIAC{b c d}{R. T. Desai}

\authorIAC{a}{F. J. Zorto-Aguilera}

\authorIAC{c e}{Z. Zhang}

\authorIAC{f}{Y. Miyake}

\affiliation{a}{Instituto de Investigación de Ciencias Aplicadas y Tecnológicas (IICAT), UNAH, Honduras}
\affiliation{b}{Centre for Fusion, Space and Astrophysics, University of Warwick, UK}
\affiliation{c}{Blackett Laboratory, Imperial College London, London, UK}
\affiliation{d}{British Antarctic Survey, Cambridge, UK}
\affiliation{e}{Stony Brook University, New York, USA}
\affiliation{f}{Kobe University, Kobe, Japan}
\email{raphael.bertrand-delgado@warwick.ac.uk}

\date{}

\begin{document}

\maketitle

\begin{abstract}
The advent of the modular CubeSat satellite architecture has heralded a revolution in satellite missions, drastically lowering the technical and financial barriers to space. As a result, over 600 CubeSat missions are due to launch in 2023 with various scientific and technology-focused applications. Surface charging resulting from energetic electron poses a direct risk to satellites in space, causing electric arcing and breakdowns. This risk is exacerbated for small technology demonstration CubeSats that are less resilient than larger satellites. An upcoming CubeSat launch of significance is the first CubeSat project originating from Honduras, the Morazán satellite (MRZ-SAT), due to launch in 2024. This will carry earth observational payloads to detect natural disasters, such as floods and landslides, which preferentially affect Central America and aims to build the first disaster forecasting capabilities for remote Central American regions. 
In this study we conduct simulations using the Electro-Magnetic Spacecraft Environment Simulator code to study absolute and differential charging of the MRZ-SAT cube-sat in Low Earth Orbit (LEO). The MRZ-SAT hosts four antennas extending from four sides of the spacecraft, an architecture which lends itself well to studying and understanding differential charging in LEO.  
The MRZ-SAT was first simulated in a typical benign ionospheric plasma environment. Here the antenna located in the ambient plasma wake displayed the maximum charging up to --0.9 V, 0.24 V biased to the main cube. An energetic electron population was then included and the wake antenna subsequently charged to greater values of --2.73 V, now 1.56 V biased to the main cube. The anisotropy of the energetic electrons was then varied, and this differential charging trend appeared exacerbated with anisotropies of 0.5 to 0.05 inducing absolute wake antenna voltages up to --4.5 V and differential voltage biases 50 and 100 \% greater than when an isotropic population was considered. This study highlights the importance of electron anisotropy in LEO to surface charging and identifies this property in the energetic electron distribution functions as inducing potentially greater risks to satellites of electrical arcing and breakdown.
\end{abstract}

\begin{multicols}{2}
\section*{Acronyms/Abbreviations}
\begin{tabular}{l m{5cm}}
    \textbf{MRZ-SAT} & Morazán satellite \\
    \textbf{UNAH} & Universidad Nacional Autónoma de Honduras \\
    \textbf{LEO} & Low Earth Orbit \\
    \textbf{MEO} & Middle Earth Orbit \\
    \textbf{GEO} & Geosynchronous Earth Orbit \\
    \textbf{DMSP} & Defense Meteorological Satellite Program \\
    \textbf{ERS} & European Remote-Sensing Satellite \\
    \textbf{EMSES} & Electro-Magnetic Spacecraft Environment Simulator \\
    \textbf{PIC} & Particle-In-Cell \\
    \textbf{HPC} & High Performance Computer \\
\end{tabular}
\section{Introduction}
In 2024, Honduras will launch the nation's first satellite, the Morazán Satellite (MRZ-SAT), to low earth orbit (LEO) \cite{monge_morazan_2019}.
The MRZ-SAT project is led by the University National Autonomy Honduras (UNAH) in collaboration with the Universidad de Costa Rica (UCR), the Universidad San Carlos de Guatemala (USAC), and Kyushu Institute of Technology, Japan.
The Morazán Project's mission has three principal aims. The first is to demonstrate a space-based early disaster warning system, informing about potential hydro-meteorological hazards, such as floods and landslides in remote and threatened areas of Honduras, Guatemala and Costa Rica. The second is to provide means of communication with the affected populations living in areas that would have possibly lost their telecommunication infrastructures during meteorological incidents \cite{fonseca_becker_morazan_2022}. Thirdly, the MRZ-SAT has an important educational purpose. Through its design and build, and subsequent data returned from on-board cameras, the mission aims to provide scientific learning tools for elementary, high-school, and university students to inspire the next generation of Honduran engineers and scientists \cite{mejuto_paving_2020}. 

Satellite surface charging is caused by energetic electrons with energies of up to several keV and presents a direct risk to satellites in space causing electric arcing and break-down \cite{gussenhoven_high-level_1985}. This risk is exacerbated for small technology demonstration Cube-Sats that are less resilient than larger satellites, often using commercial off the shelf components. Compared to geostationary earth orbit (GEO) where satellites encounter the high-temperature plasma sheet on the night-side \cite{deforest_spacecraft_1972, choi_analysis_2011}, the plasma environment in LEO is colder, and spacecraft charging is not typically considered as great a risk as at GEO \cite{ferguson_feasibility_2014, belov_relation_2005}. Nonetheless, significant charging has been observed in polar regions; for example, a potential of --2000 V was regularly measured on the satellite DMSP F12 \cite{anderson_characteristics_2012}. Another example is the complete loss of the ERS-1 satellite in March 2000 and the ASCA satellite in July of the same year, both satellites having respectively an altitude of 772 km and 570 km \cite{noauthor_esa_nodate,noauthor_heasarc_nodate}. More recently, the Jason-3 satellite has reported charging events of up to --2000 V negative for a variety of geomagnetic conditions and notably down to 60$^\circ$ latitudes \cite{f_enengl_characterization_2022}.  %

 Energetic electrons at LEO are produced when enhanced solar wind conditions drive magnetic reconnection and substorms in the magnetotail. Energetic electrons subsequently precipitate into the polar atmosphere and produce the polar aurora and associated Region 1 current system \cite{birkeland_norwegian_1908, birkeland_norwegian_1913}. Following the impact of solar storms \cite{Koehn22}, the magnetosphere can become highly compressed, bringing the night-side reconnection X-line closer to Earth \cite{Desai21} and expanding the auroral oval to lower latitudes \cite{carbary_kp_2005, hardy_statistical_1985}. A consequent build up of the partial night-side and global ring current results in the formation of further Region 2 current systems at even lower latitudes \cite{nesse_energetic_2023}. Energetic electrons in LEO derive from kinetic plasma instabilities which scatter particles into the loss cone and from the auroral acceleration region (AAR) \cite{marghitu_anisotropy_2006}.

A spacecraft immersed within a plasma will gain a net charge despite the condition of quasi-neutrality. The electrons are much lighter than the ions and move at a higher velocity and therefore preferentially stick to the spacecraft surfaces causing a net negative charge within a distance of the Debye length \cite{whipple_potentials_1981}. Electron emission processes, due to sunlight or particle impact, can conversely cause the surface to emit electrons and the net charge to move to positive potentials. Spacecraft charging risks are exacerbated when different parts of the spacecraft charge faster than the charge can equalise, thus causing differential charging \cite{garrett_charging_1981}. A difference of potential can lead to a discharge by electric arcs, a hazard that can partially or completely damage the satellite.

Simulating a satellite embedded inside plasmas can help to explain and mitigate adverse charging behaviour. In this article is a description of a self-consistent three-dimensional study of the MRZ-SAT in LEO. The simulation method is first discussed in Section \ref{sec:method}, along with the satellite architecture, environmental conditions and inclusion of energetic electrons. Section \ref{sec:background} first analyses the electric charging of the MRZ-SAT in conditions representative of the ionospheric plasma at 400 km of altitude \cite{pignalberi_three-dimensional_2019}. Energetic electrons are then implemented in the simulations in Section \ref{sec:he}, in order to study their impact on the net electric charge between different parts of the spacecraft. Subsection \ref{sec:anisotropy} studies the effects of electron anisotropy on net and differential charging.

\section{Method}
\label{sec:method}
\subsection{EMSES simulations technique}
This study utilises the three-dimensional electromagnetic spacecraft environment simulator (EMSES) code, which simulates spacecraft-plasma interactions using the Particle-In-Cell (PIC) method \cite{miyake_new_2009}. The PIC approximation corresponds to replacing the high amount of electrons and ions with charged macro-particles to reproduce a continuous phase space. This code has successfully been applied to study spacecraft charging at Earth \cite{miyake_new_2009, miyake_electron_2020}, and Saturn \cite{zhang_particle--cell_2021, zhang_simulating_2023,ZhangURSI}.

The plasma flow is injected as a drifting Maxwellian and exits through an outflow boundary condition. Further boundaries are periodic, allowing a continuous space orthogonal to the plasma flow. Each plasma species has mass and charge normalised to the proton scale with a real ion-to-electron mass ratio. The spacecraft body can be composed of multiple structures, either perfectly conducting or electrically insulated from each other. The charge accumulation caused by impinging particles is redistributed over its whole surface in order to maintain an equipotential distribution via the capacity matrix method. The charge density is used to solve Poisson’s equation for the electrostatic potential. 
The simulated domain is a 3-D grid, with a spacing chosen to resolve the Debye length scale. 

\subsection{Morazán Satellite}
\begin{figure}[H]
    \centering
    \includegraphics[width = 1\linewidth]{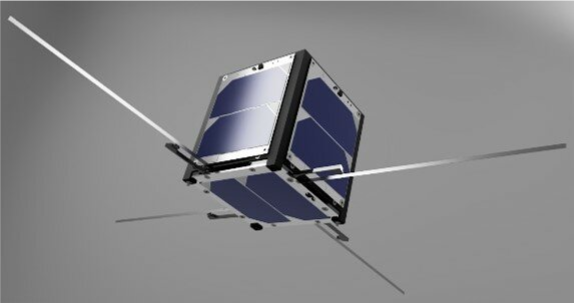}
    \caption{CAD model of the MRZ-SAT structure.}
    \label{fig:MRZ-SAT_structure}
\end{figure}

\begin{figure}[H]
    \centering
    \includegraphics[width = 1\linewidth]{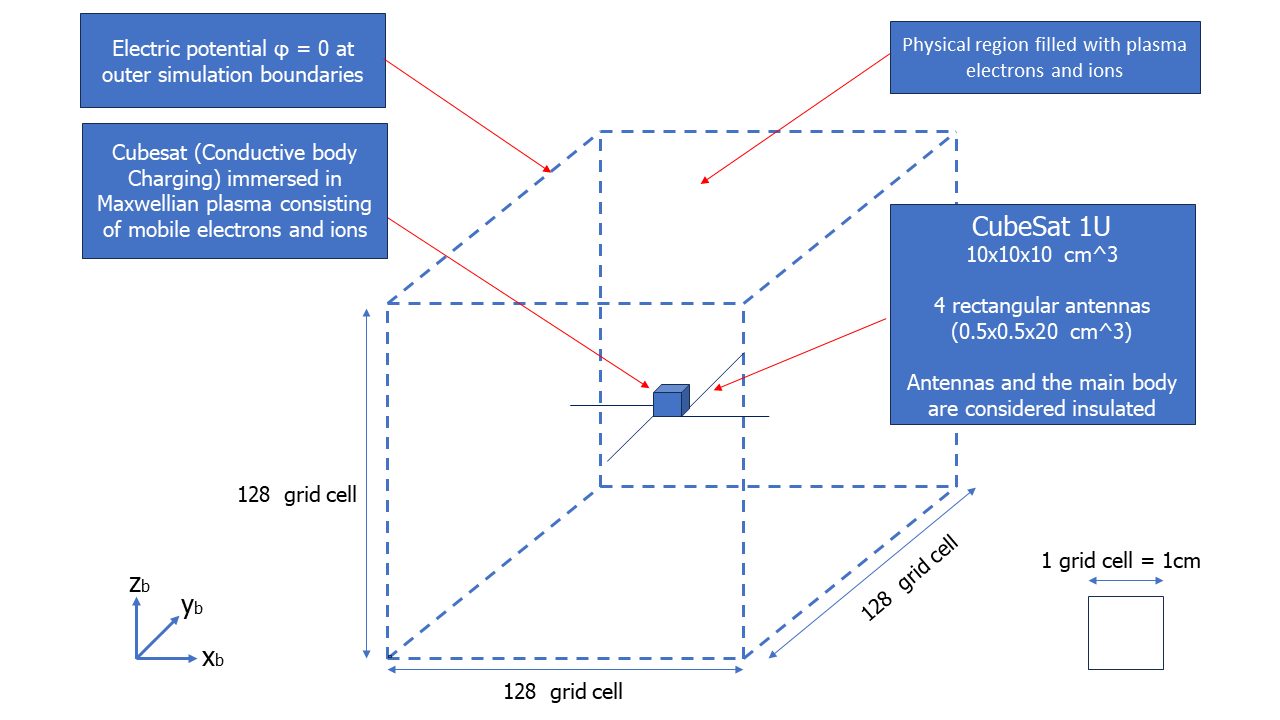}
    \caption{Representatin of the MRZ-SAT in the EMSES domain. A subset surrounding the spacecraft is shown in the subsequent simulationr results.}
    \label{fig:Emses model}
\end{figure}

 Figure \ref{fig:MRZ-SAT_structure} shows the 1U MRZ-SAT as a main cube of dimensions $10\ \times \ 10 \ \times \ 10\ {\rm cm^3}$ \cite{hipp_morazan_2022} and Figure \ref{fig:Emses model} shows this represented within the EMSES simulation domain that has dimensions of $128 \times 128 \times 128$ grid cells, with a grid width of $1$ cm, as shown in Figure \ref{fig:Emses model}. Four rectangular antennas extend outward from each edge with dimensions $0.5\ \times 0.5\ \times 20\ {\rm cm^3}$. The four antennas are all located on the $({\rm {\bf X_B},\ {\bf Y_B}})$ plane set on the z upper part of the cube, two along ${\rm {\bf X_B}}$, amongst the two, one is on the lower half of the plane (between $x = 0$ and $x = 64$). A $64 \times 64 \times 64$  cm$^3$ subset of the simulation domain is used for visualisation and the antenna are referred to as x- or y-negative and x- or y-positive in accordance with the direction they extend from the main spacecraft in this coordinate system.
 In this study, the four antennas and main body are considered to be electrically insulated from one another to examine differential charging phenomena. While in reality charge can flow between the different structures, this assumption will reproduce the underlying behaviour of potentially hazardous differential charging due to non-conducting elements involved in cube-sat designs, and particularly during rapid onset events where charge accumulates faster than it can equilibriate throughout the spacecraft and leak away into the ambient plasma.   
 
\subsection{Ionospheric plasma}
\label{sec:atmos_plasma}
The MRZ-SAT will be released from the International Space Station (ISS) and will therefore follow a similar orbit at an altitude of 400 km and an orbital velocity of 7.9  km/s \cite{hipp_morazan_2022}. At this altitude, positively charged particles from the ionosphere are mainly ions \O \cite{pignalberi_three-dimensional_2019}, therefore in this study, these are considered the only injected positive ions. Both species, ions and electrons, are considered to have the same temperature: $T_e = T_i$. 
The ionospheric plasma is considered coming along ${\rm {\bf X_B}}$ from negative to positive with parameters as shown in Table \ref{tab:background_plasma}.

\begin{table*}[ht]
    \centering
    \caption{Ionospheric plasma and system parameters.}
    \begin{tabular}{lc}
    \specialrule{.15em}{.2em}{.2em} 
       Electron density, $n_{e,0}$ [${\rm cm^{-3}}$] & $2.50\times 10^5$\\
       Electron temperature, $T_e$ [K] & $2000$\\
       Ion temperature, $T_i$ [K] & $2000$\\
       Flow speed, ${\bf v_{flow}}$ [m/s] & $7.90\times10^3\ {\bf \hat{x}}$ \\
       Magnetic field [nT] & 36.99, 0, -22.97 \\
       Energetic electron flux [${\rm keV/cm^2/s/sr}$] & $1.56\times10^{9}$   \\
       Energetic electron energy [keV] & $1-1.09$ \\
       \specialrule{.15em}{.2em}{.2em} 
       Grid size [cm] & $0.5-1$ \\
       Time step [s] & $3.34\times10^{-10}$ \\
       Particles per cell & 41 943 040 \\
       Domain size [${\rm cm^{3}}$] &  $128\times128\times128$ \\
       \specialrule{.15em}{.2em}{.2em} 
    \end{tabular}
    \label{tab:background_plasma}
\end{table*}


\subsection{Magnetic field and energetic electrons}
\label{sec:mag-field_ae}
\begin{figure}[H]
    \centering
    \includegraphics[width = 1\linewidth]{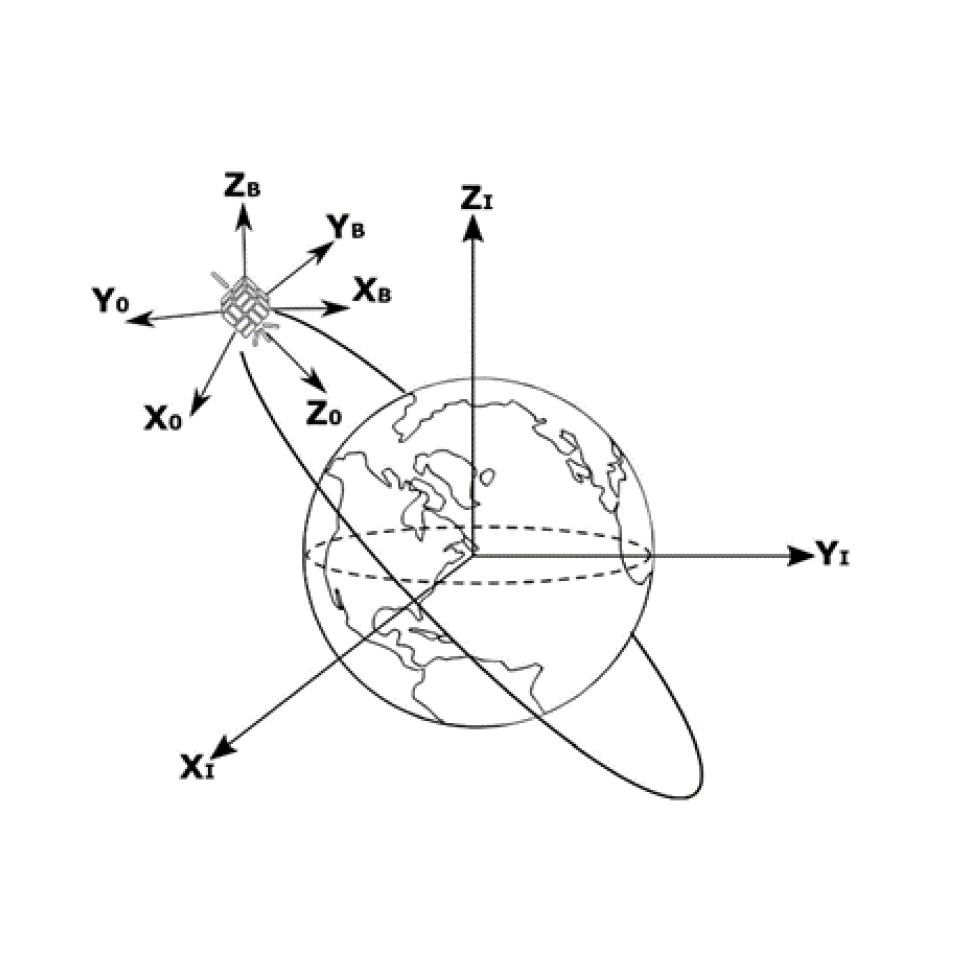}
    \caption{MRZ-SAT reference coordinate system.}
    \label{fig:ref_systems}
\end{figure}
Figure \ref{fig:ref_systems} presents the different reference systems used. ${\rm ({\bf X_I,\ Y_I,\ Z_I})}$ characterises the Geocentric Reference System. ${\rm ({\bf X_0,\ Y_0,\ Z_0})}$ is the Orbital Reference System, the origin situated at the cubesat’s centre of  mass,  ${\rm {\bf X_0}}$  points  in  the  same  direction  as  the cubesat’s  linear  velocity  vector  (Ram  direction), tangent to the orbit it describes, and ${\rm {\bf Z_0}}$  always points in the direction of the Earth’s centre. In this study, the used system will generally be the Moving Reference System ${\rm ({\bf X_B,\ Y_B,\ Z_B})}$ characterised by being free to translate or rotate with the cubesat. Particularly, in the simulation the Moving Reference System will be taken as ${\rm {\bf X_B}} = {\rm {\bf X_0}}$ and ${\rm {\bf Z_B}} = {\rm -{\bf Z_0}}$. \\
Following the ISS orbit, the cubesat maximum latitude will be 51.65° \cite{hipp_morazan_2022}. Most of the energetic electron precipitations are detected at a latitudes at or greater than 55$^\circ$ \cite{enengl_characterization_2022,nesse_energetic_2023}. However, during large solar storms, ionospheric current systems and energetic electrons can reach latitudes lower than 50$^\circ$ \cite{carbary_kp_2005, hardy_statistical_1985}. Indeed, the ASCA satellite, located at an altitude of 570 km, was completely lost on $15^{{\rm}}$ July 2000 after a geomagnetic storm where the Kp index reached 9 \cite{noauthor_heasarc_nodate}. 
For the MRZ-SAT at peak latitude, we use a magnetic field under the assumption of a dipole with moment, 7.94 A$\cdot {\rm m^2}$, with strength of $43.14\ {\rm \mu T}$ at MRZ-SAT, and oriented 58.14$^\circ$ relative to ${\rm {\bf Z_I}}$. \\
In this study the MRZ-SAT is considered on the night-side of Earth with zero solar illumination, thus without photoelectron emission. A range of energetic electrons are considered, from 0.1 to 10 keV, for fluxes of $10^{10} - 10^{12}\ {\rm /cm^2/s}$. Electron anisotropy is well known to affect spacecraft charging at GEO and Middle Earth Orbit (MEO) \cite{kletzing_auroral-plasma_1999, olsson_recent_2003}, but this has yet not been identified or studied in LEO. Electron anisotropy will be represented in the simulation by partitioning the electron energy between the thermal energy for the case of an isotopic distribution and the drift velocity for the case of an electron beam, and variations in between. In each scenario the simulated energy flux will be equal to $1.56\times10^{9}\ {\rm keV / cm^2/ s / sr}$.

\section{Plasma interaction}
\label{sec:background}
\begin{figure}[H]
    \centering
        \includegraphics[width = 1.0 \linewidth]{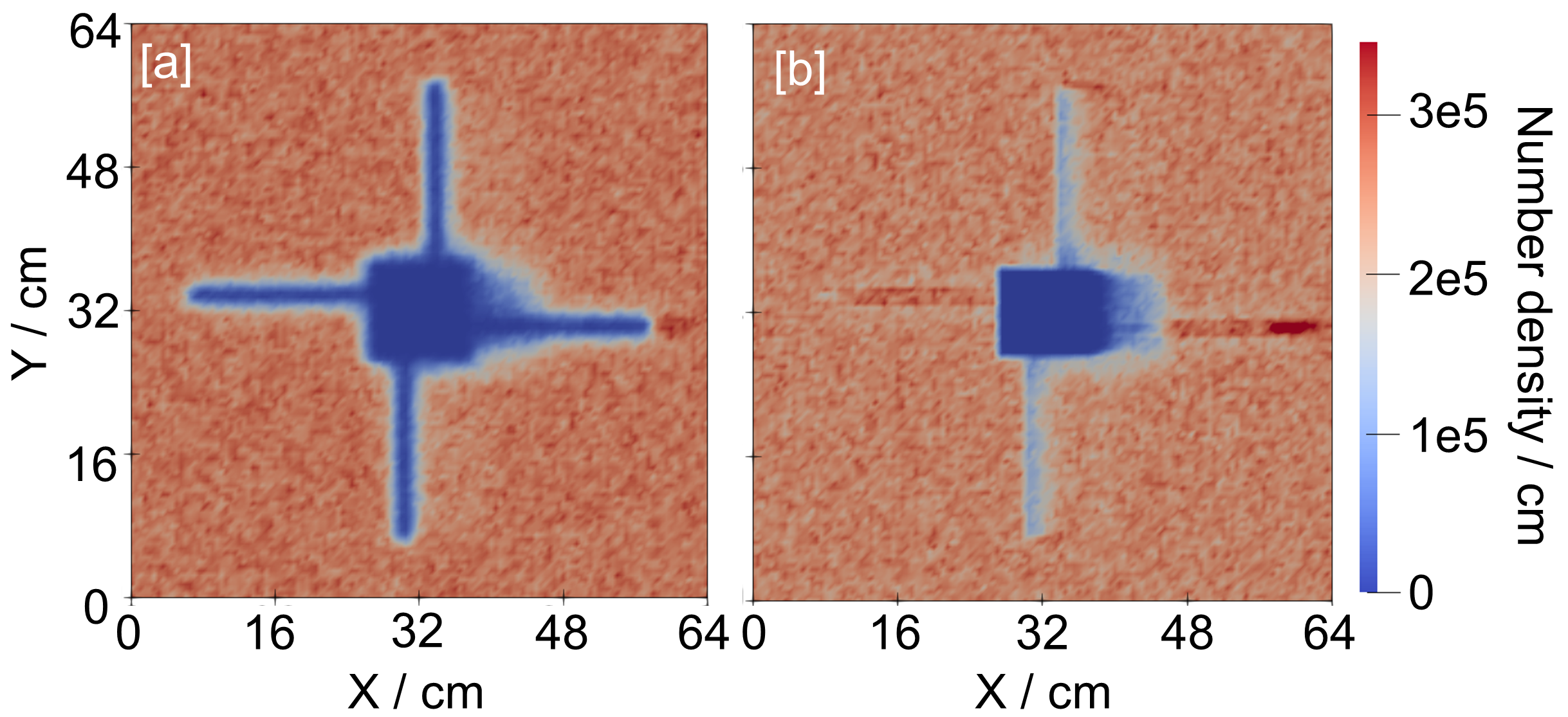}
        \caption{Subset of the domain showing the simulated ambient plasma interaction. The electron (left) and ion (right) densities are shown in the spacecraft frame with the plasma entering from the left.}
        \label{fig:paraview_ref}
\end{figure}

The Morazán Satellite is first simulated in a typical plasma environment without energetic electrons to understand the ambient plasma interaction. The ionospheric plasma parameters are given in Section \ref{sec:atmos_plasma}. \\
Figure \ref{fig:paraview_ref} presents two slices on the $({\rm {\bf X_B},\ {\bf Y_B}},\ Z_B=72)$ plane and Table \ref{tab:pot_background} the final potentials reached by each spacecraft structure and their difference relative to the main cube. \\
The negative spacecraft potentials cause the spacecraft and antennas to appear as regions devoid of electrons but with enhanced ion fluxes. The enhanced ions fluxes are due to the antenna physical domain being implemented sub-grid and are therefore not visible as empty regions of plasma. A low density wake is present behind the satellite with the ion wake appearing larger than the electron wake due to their larger inertia and therefore slower refilling of the wake. \\
Table \ref{tab:pot_background} shows the potential of the different spacecraft structures. The potential varies between --0.61 and --0.9 V with the x-positive antenna having a lower potential than the rest of the conducting bodies. This can be explained by the wake visible behind the satellite displaying a lower plasma density surrounding the antenna and the latter therefore experiencing a higher ratio of ionospheric electrons to \O ions.
The difference with the main cube is also shown, which indicates that the highest difference reaches $0.24$ V. 
\begin{table}[H]
    \centering
    \caption{Final energy potential for each component of the cubesat and their final bias relative to the main cube for the satellite for the scenarios without energetic electrons.}
    \begin{tabular}{lcc}
    \specialrule{.15em}{.2em}{.2em}
        Component & Potential [V] & Bias [V] \\
        Cube & -0.66 & 0\\
        Antenna x-negative & -0.73 & 0.17\\
        Antenna x-positive & -0.90 & 0.24\\
        Antenna y-negative & -0.70 & 0.04\\
        Antenna y-positive & -0.61 & -0.05\\
        \specialrule{.15em}{.2em}{.2em}
    \end{tabular}
    \label{tab:pot_background}
\end{table}

\section{Energetic electrons}

\label{sec:he}

\subsection{Energetic electrons}
\label{sec:he_mid-flux}
\begin{figure}[H]
    \centering
        \includegraphics[width = 1.0 \linewidth]{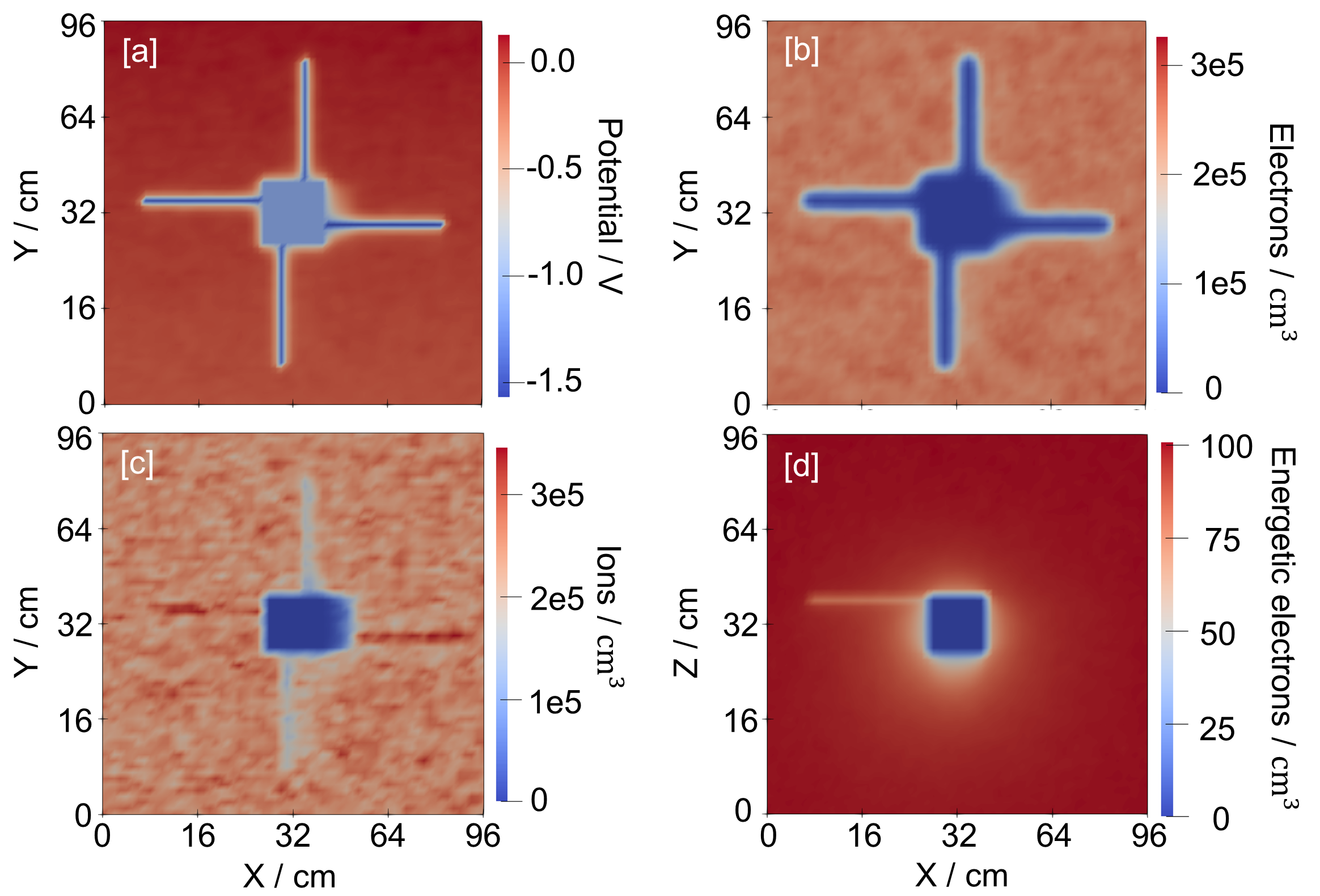}
        \label{fig:paraview_pot_90-10}
    \caption{Slices showing the potential (top left, a), electron (top right, b), ion (bottom left, c), and energetic electron (bottom right, d) densities at a final sate, with an energy flux of $1.56\times10^{9}\ {\rm keV / cm^2/ s / sr}$.}
    \label{fig:paraview_90-10}
\end{figure}

To simulate energetic electron fluxes that the MRZ-SAT may experience, an energy flux of $1.56\times10^{9}\ {\rm keV / cm^2/ s / sr}$ was simulated, based upon Jason-3 and ASCA observations \cite{carbary_kp_2005, enengl_characterization_2022}. Due to electrons moving significantly faster than the spacecraft, the electron energy is considered primarily as a thermal energy of 1 keV. Their drift velocity is of a similar order of magnitude to the spacecraft velocity and they are therefore intended to represent an isotropic flow. \\
Figure \ref{fig:paraview_90-10} and Table \ref{tab:pot_1kev_90-10} show the simulation results where energetic electrons are included. Compared to the previous charging in Section \ref{sec:background}, energetic electrons induce a greater potential which results in further deflection of ionospheric electrons from the spacecraft. 
The high flux of energetic electrons along the magnetic field direction in Figure \ref{fig:paraview_90-10}d, produces a slight distortion in the energetic electron distribution towards the bottom left-hand corner of the simulation box. \\
The energetic electrons causes the MRZ-SAT to charge to --1.19. Figure \ref{fig:paraview_90-10}a shows the antenna x-positive charges the most. Figure \ref{fig:paraview_90-10}d shows a slice in the $({\rm {\bf X_B},\ {\bf Z_B}})$ plane where $Y_B= 68$ and therefore only shows one antenna.
Table \ref{tab:pot_1kev_90-10} shows that the x-positive antenna has a final potential of $-2.73$ V, with a difference of potential of $1.56$ V with the main cube, while the three others have a differences of --0.46 to --0.67 V. \\
The energetic electrons therefore play an important role in spacecraft charging, explaining the enhanced difference between the x-positive antenna and the rest of the satellite compared with the case without energetic electrons.
\begin{table}[H]
    \centering
    \caption{Final energy potential for each component of the cubesat and their final bias relative to the main cube with energetic electrons at a thermal energy of 1 keV and a kinetic drift energy of 0.1 keV.}
    \begin{tabular}{lcc}
    \specialrule{.15em}{.2em}{.2em}
        Component & Potential [V] & Bias [V] \\
        Cube & -1.19 & 0 \\
        Antenna x-negative & -1.86 & 0.67 \\
        Antenna x-positive & -2.73 & 1.56\\
        Antenna y-negative & -1.67 & 0.49\\
        Antenna y-positive & -1.64 & 0.46\\
        \specialrule{.15em}{.2em}{.2em}
    \end{tabular}
    \label{tab:pot_1kev_90-10}
\end{table}
\subsection{Anisotropic Energetic Electrons}
\label{sec:anisotropy}
The electron anisotropy is defined by the ratio T$_\perp/T_\parallel$ relative to the magnetic field. Electron showers have been measured with variable anisotropies, depending on the energetic electron's energy and the Kp index \cite{marghitu_anisotropy_2006, kletzing_auroral-plasma_1999}. Olsson and Janhunen \cite{olsson_recent_2003} especially study the characteristics of `middle-energy' energetic electrons, in a range between 100 and 1000 eV.
 
\subsubsection{Anisotropy of 0.5}
\label{sec:50-50case}
\begin{figure}[H]
    \centering
        \includegraphics[width = 1.0 \linewidth]{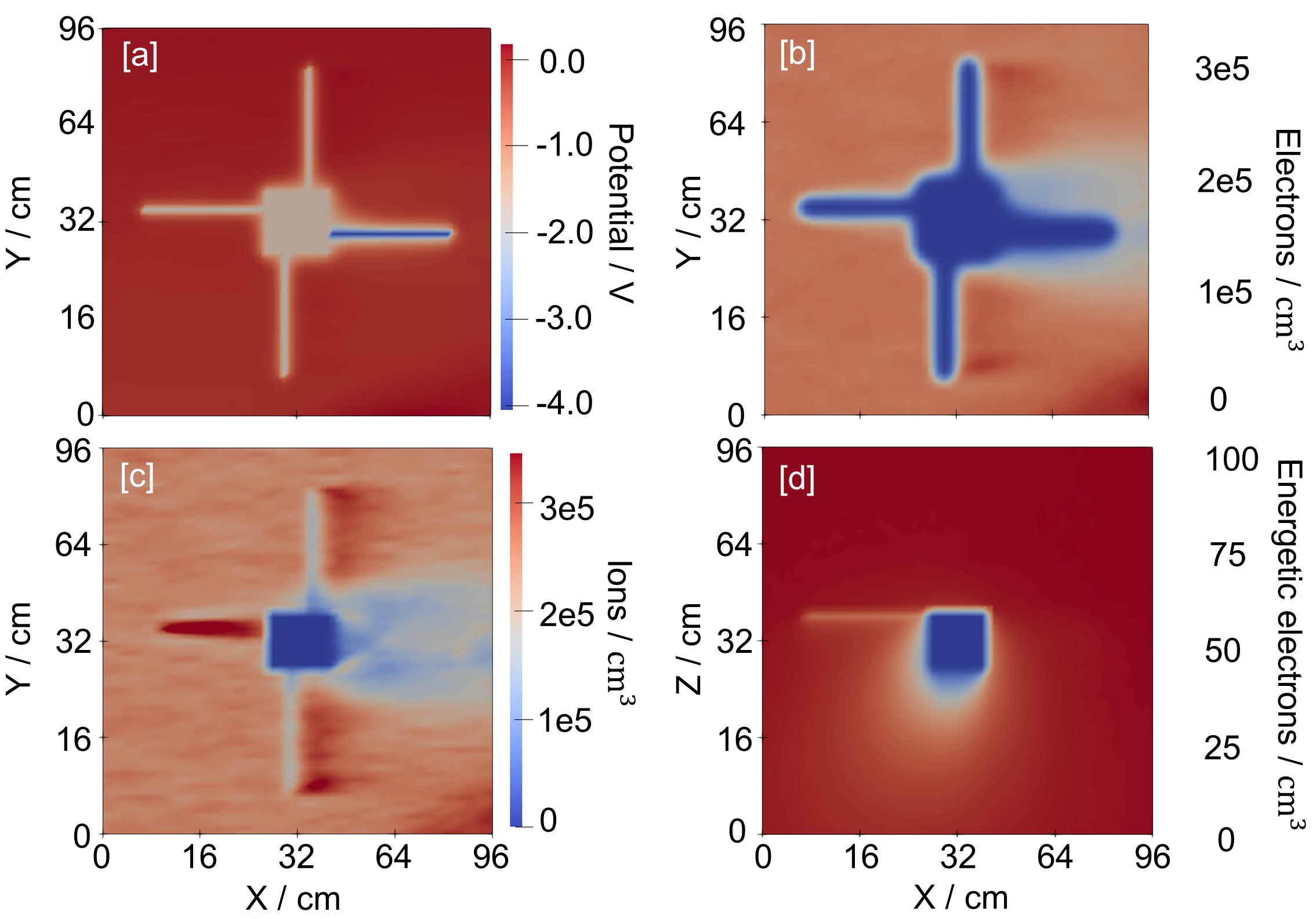}
    \caption{Slices showing the potential (top left, a), electron (top right, b), ion (bottom left, c) and energetic electron (bottom right, d) densities at a final sate, with an anisotropy of 0.5.}
    \label{fig:paraview_50-50}
\end{figure}

The energetic electrons are first considered with an initial anisotropy of 0.5 \cite{marghitu_anisotropy_2006}. Figure \ref{fig:paraview_50-50} and and Table \ref{tab:pot_1kev_50-50} present the simulation results where the the energetic electrons are also implemented with a the same total energy of 1 keV and flux as in Section \ref{sec:he_mid-flux}. However here, the energy is equally divided between the drift velocity and the thermal energy. Therefore the parallel kinetic energy equals 500 eV, resulting in a field-aligned drift velocity of $1.33\times10^7\ {\rm m/s}$. 
\begin{table}[H]
    \centering
    \caption{Final energy potential for each component of the cubesat and their final bias relative to the main cube for energetic electrons at a thermal energy of 0.5 keV and a kinetic energy of 0.5 keV.}
    \begin{tabular}{lcc}
    \specialrule{.15em}{.2em}{.2em}
        Component & Potential [V] & Bias [V] \\
        Cube & -1.73 & 0 \\
        Antenna x-negative & -2.01 & 0.28\\
        Antenna x-positive & -4.54 & 2.81\\
        Antenna y-negative & -2.07 & 0.34\\
        Antenna y-positive & -1.86 & 0.13\\
        \specialrule{.15em}{.2em}{.2em}
    \end{tabular}
    \label{tab:pot_1kev_50-50}
\end{table}
Once again, it is the x-positive antenna that reaches the most negative potential. Table \ref{tab:pot_1kev_50-50} indicates a more negative final potential of --4.54 V and a final difference with the main body of 2.81 V, a difference relative to the main cube higher than for an isotropic electron distribution examined Section \ref{sec:he_mid-flux}. The other antennas, however display differential biases of 0.13--0.34 which are lower than when isotropic electrons were considered.\\
Figure \ref{fig:paraview_50-50} shows the plasma interaction and notable differences to Figure \ref{fig:paraview_90-10}. Figures \ref{fig:paraview_50-50}a and \ref{fig:paraview_50-50}b show how the enhanced charging of the x-positive antenna deflects the ambient electrons to a greater extent causing slower refilling of the wake and therefore a much larger electron wake. The ion wake in Figure \ref{fig:paraview_50-50}c is also larger and highlights how the ions are coupled to the electrons via ambi-polar electric fields. The energetic electron wake oriented along the magnetic field is also understandably more pronounced due to the greater drift velocity of this population.

\subsubsection{Anisotropy of 0.05}
\label{sec:beam_case}
\begin{figure}[H]
    \centering
        \includegraphics[width = \linewidth]{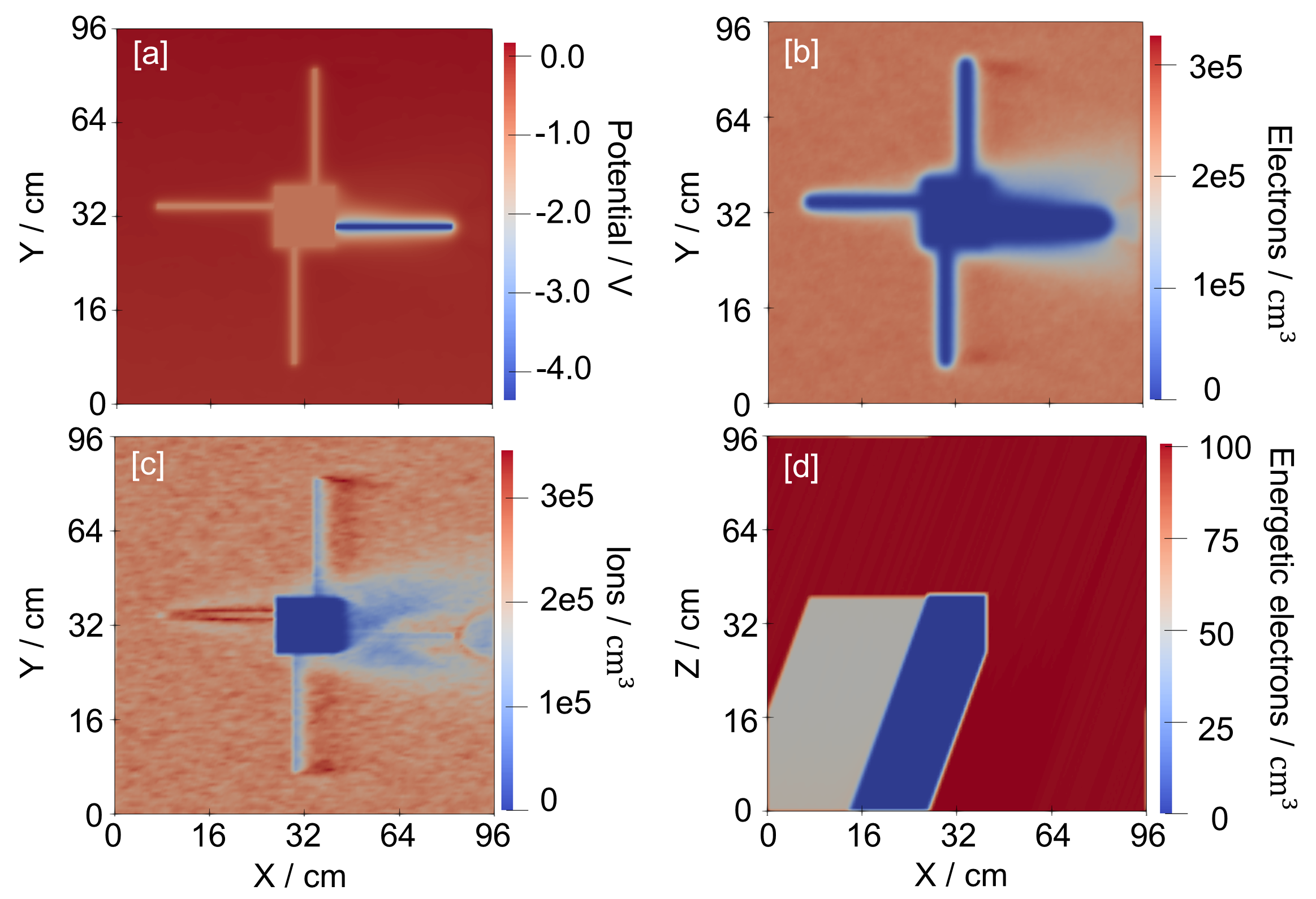}
    \caption{Slices showing the potential (top left, a), electron (top right, b), ion (bottom left, c) and energetic electron (bottom right, d) at a final sate, and an anisotropy of 1/20.}
    \label{fig:paraview_beam}
\end{figure}
The second study case of anisotropy looks to examine the limiting scenario of a pure electron beam and thus simulates an energetic population with a kinetic energy of 1 keV and a thermal perpendicular energy of just 0.1 keV.
 Figure \ref{fig:paraview_beam} and Table \ref{tab:pot_1kev_beam} shows the simulation and potential charging results.
\begin{table}[H]
    \centering
    \caption{Final energy potential for each component of the cubesat and their final bias relative to the main cube for energetic electrons at a thermal energy inferior to 0.1 keV and a kinetic energy of 1 keV.}
    \begin{tabular}{lcc}
    \specialrule{.15em}{.2em}{.2em}
        Component & Potential [V] & Bias [V] \\
        Cube & -0.91 & 0 \\
        Antenna x-negative & -1.10 & 0.19 \\
        Antenna x-positive & -4.41 & 3.5\\
        Antenna y-negative & -1.06 & 0.15\\
        Antenna y-positive & -0.99 & 0.08\\
        \specialrule{.15em}{.2em}{.2em}
    \end{tabular}
    \label{tab:pot_1kev_beam}
\end{table}

The most evident difference in the plasma interactions is the elongated sharp energetic electron wake in Figure  \ref{fig:paraview_beam}d. The ion wake in Figure \ref{fig:paraview_beam}c appears similar to in Figure \ref{fig:paraview_50-50}d but the electron void surrounding the x-positive antenna in Figure \ref{fig:paraview_beam}b is larger than the case shown in Figure \ref{fig:paraview_50-50}b. 

The x-positive antenna reaches a negative potential of --4.41 V, an absolute potential comparable to the previous case displayed in Section \ref{sec:50-50case}. The other antennas and main cube, however, reach less negative values of --0.91 V to --1.06 V, nearly half as negative as in the case where the anisotropy was 0.5. This is attributed to the directional flow of the energetic electron beam reducing cross-sectional antenna and cube area visible to the energetic electrons. This lower charging of the main cube, however, interestingly causes the differential charging between the the x-positive antenna and main cube to be 25 \% greater reaching 3.5 V. 

This dynamic of differential charging shows how the altered wake charging scales differently to the main body and results in greater differential charging.

\section*{Conclusion}
In this study we have examined the absolute and differential charging of the Morazán MRZ-SAT satellite in its target orbit of 400 km and highest latitude of 51.65$^\circ$. The MRZ-SAT was first simulated subjected to the ambient thermal electron and \O ion plasma conditions. This resulted in small amount of absolute and differential charging, with the x-positive antenna charging greater than other antennas due to the shadowing of reduced wake densities.

At the latitudes considered, spacecraft are not typically be exposed energetic electron but, during geomagnetic storms, current systems can close at lower latitudes and combined with enhanced magnetospheric and AAR wave activity, can produce energetic electrons in LEO at these latitudes \cite{carbary_kp_2005, hardy_statistical_1985}. Field-aligned energetic electrons of 1 keV were first injected as a near-isotopic distribution and then with varying degrees of anisotropy relative to the magnetic field, firstly of 0.5 and then an order of magnitude lower of 0.05.
The injection of this additional energetic electron population along a different axis with a near isotropic flow induced a slight distortion to the density distribution with slightly reduced densities appearing along the magnetic field. This effect was exacerbated for anisotropic energetic electron distributions with greater field-aligned drift velocities and a clear double wake structure was produced. The anisotropic electrons were subsequently found to produce lower overall charging of the main cube and of three of the antennas but the wake antenna interestingly exhibited greater charging and therefore the potential bias to the main cube was increased.  

This study has identified that electron anisotropy can potentially induce greater differential charging to a satellite exposed to energetic electrons in LEO. This highlights a further parameter of interest when examining surface charging risks to satellites in LEO, in addition to their energies and flux.\\
Further studies could deepen the understanding of the effects of plasma anisotropy on surface charging and directions to be considered include accounting for secondary electron emissions from the spacecraft \cite{balcon_secondary_2012} and ionospheric density depletions during geomagnetic storms, both phenomena are often coincident with energetic electrons \cite{crowley_global_2006, dent_plasmaspheric_2006}, as well as different energetic electron energies beyond those initially considered herein .

\section*{Acknowledgements}
R.B.D. acknowledges the financial support from the UNAH and the University of Warwick. 

R.T.D. acknowledges an STFC Ernest Rutherford Fellowship ST/W004801/1. This work used the High Performance Computing Service from the University of Warwick.

\section*{Data Availability}

\bibliographystyle{ieeetr}
\bibliography{main}

    
 
  

\end{multicols}

\end{document}